\documentclass[12pt]{article}
\usepackage{graphicx,epsfig}
\usepackage[nospace]{cite}

\newcommand{\er}{{\bf e}_{r}}
\newcommand{\ep}{{\bf e}_\varphi}
\newcommand{\et}{{\bf e}_\theta}
\newcommand{\hh}{{\bf H}}
\newcommand{\ee}{{\bf E}}
\newcommand{\ww}{{\bf W}}

\newcommand{\rr}{{\bf r}}

\newcommand{\lom}{{\bf L}}

\newcommand{\rmi}{{\rm i}}

\begin{document}

\title{Electrodynamic spherical harmonic}

\author{A V Novitsky\\
Department of Theoretical Physics, Belarusian State University, \\
Nezavisimosti~Avenue~4, 220050 Minsk, Belarus \\
andrey.novitsky@tut.by}

\date{}


 \maketitle

\begin{abstract}
Electrodynamic spherical harmonic is a second rank tensor in
three-dimensional space. It allows to separate the radial and angle
variables in vector solutions of Maxwell's equations. Using the
orthonormalization for electrodynamic spherical harmonic, a boundary
problem on a sphere can be easily solved.
\end{abstract}

\section{Introduction}

In this paper we introduce new function --- electrodynamic spherical
harmonic. It is represented as a second rank tensor in
three-dimensional space. But the function differs from tensor
spherical harmonic \cite{Newman,James,Sandberg,Winter,Kostelec}. By
its definition the electrodynamic spherical function possesses a
number of properties of usual scalar and vector harmonics and
includes them as component parts. Why {\it electrodynamic} spherical
harmonic? The word ``electrodynamic'' implies that the function is
applied for solution of {\it vector} field problems. We use it in
electrodynamics, however one can apply the function for description
of spin fields in quantum field theory.

Electrodynamic spherical harmonic is not a simple designation of the
well-known functions. It satisfies the Maxwell equations and
describes the angular dependence of vector fields. The introduced
function separates the variables (radial coordinate and angles) in
the fields. Moreover, the notation of the fields in terms of
electrodynamic spherical harmonic noticeably simplifies the solution
of a boundary problem on spherical interface. Just application of
the ortonormalization condition allows to find the coefficients of
spherical function expansion (e.g. scattering field amplitudes).

\section{Scalar and vector spherical harmonics}

Hamilton's operator $\nabla$ in three-dimensional space contains the
derivatives on three coordinates. In spherical coordinates ($r$,
$\theta$, $\varphi$) the derivatives on radial coordinate and angles
can be separated. This is achieved by representing the unit tensor
${\bf 1}$ in 3D space as the superposition of two projection
operators:
\[
\nabla={\bf 1} \nabla = \left(
\frac{\rr\otimes\rr}{r^2}-\frac{\rr^{\times 2}}{r^2} \right) \nabla
= \frac{1}{r^2} \rr(\rr \nabla) - \frac{1}{r^2} \rr^\times
(\rr^\times \nabla), 
\]
where $\rr\otimes\rr/r^2$ is the projector onto the direction
$\er=\rr/r$, $-\rr^{\times 2}/r^2$ is the projector onto the plane
orthogonal to the unit vector $\er$. Tensor $\rr^\times$ dual to the
vector $\rr$ gives the well-known vector product when acting on a
vector ${\bf a}$: $\rr^{\times} {\bf a}=\rr \times {\bf a}$ and
${\bf a} \rr^{\times}={\bf a} \times \rr$ \cite{Fedorov}.
Introducing the vector differential operator
\begin{equation}
\lom=\frac{1}{\rmi} \rr \times \nabla \label{L-oper}
\end{equation}
the equation (\ref{nabla1}) is rewritten as follows
\begin{equation}
\nabla=\frac{\rr}{r} \frac{\partial}{\partial r} - \frac{\rmi}{r^2}
\rr \times \lom. \label{nabla2}
\end{equation}

Vector $\lom$ is called orbital angular momentum operator in quantum
mechanics, because it is presented as vector product of radius
vector $\rr$ and momentum ${\bf p}=-\rmi \nabla$ operators. $\lom$
includes only derivatives on the angles $\theta$ and $\varphi$.
Using the definition of $\lom$ one can derive the Laplace operator
\begin{equation}
\Delta=\nabla^2=\frac{1}{r^2} \frac{\partial}{\partial r}\left( r^2
\frac{\partial}{\partial r} \right) - \frac{\lom^2}{r^2}
\label{laplace}
\end{equation}
and the following properties:
\begin{equation}
\rr \lom =0, \qquad (\lom \rr) =0, \qquad \lom^2 \lom =\lom \lom^2,
\qquad \lom \times \lom = \rmi \lom. \label{L-properties}
\end{equation}

Scalar spherical harmonic $Y_{lm} (\theta, \varphi)$ is defined as
the eigenfunction of the operator~$\lom^2$:
\begin{equation}
\lom^2 Y_{lm} = l (l+1) Y_{lm}, \label{L2}
\end{equation}
where $l$ and $m$ are integer numbers. Number $m$ is the eigenvalue
of the operator of projection of angular momentum onto the axis $z$,
$L_z$:
\begin{equation}
L_z Y_{lm} = m Y_{lm}. \label{Lz}
\end{equation}

Spherical harmonics $Y_{lm}$ are orthogonal and normalized by the
unit:
\begin{equation}
\int_0^\pi \int_0^{2 \pi} Y^\ast_{l'm'}(\theta, \varphi)
Y_{lm}(\theta, \varphi) \sin \theta {\rm d} \theta {\rm d} \varphi =
\delta_{l'l} \delta_{m'm}. \label{orthogY}
\end{equation}
where the sign $^\ast$ denotes the complex conjugate.

If we multiply equation (\ref{L2}) by the vector operator $\lom$ and
take into account the commutation of $\lom$ and $\lom^2$, we will
obtain that vector $\lom Y_{lm}$ satisfies the same equation
(\ref{L2}), too. The quantity defined as
\begin{equation}
{\bf X}_{lm}=\frac{1}{\sqrt{l (l+1)}} \lom Y_{lm}
\end{equation}
is called vector spherical harmonic. The coefficient before $\lom
Y_{lm}$ is chosen so that the orthonormalization condition is of the
form
\begin{equation}
\int_0^\pi \int_0^{2 \pi} {\bf X}_{l'm'}^\ast(\theta, \varphi) {\bf
X}_{lm}(\theta, \varphi) \sin \theta {\rm d} \theta {\rm d} \varphi
= \delta_{l'l} \delta_{m'm}. \label{orthogX}
\end{equation}

From the self-conjugacy of the angular momentum operator $\lom$ and
properties (\ref{L-properties}) the orthogonality
\begin{eqnarray}
\int_0^\pi \int_0^{2 \pi} \er ({\bf X}_{l'm'}^\ast \times {\bf
X}_{lm}) \sin \theta {\rm d} \theta {\rm d} \varphi &=& \int_0^\pi
\int_0^{2 \pi} \frac{ Y_{l'm'}^\ast \er (\lom \times \lom)
Y_{lm}}{\sqrt{l (l+1) l' (l'+1)}} \sin \theta {\rm d} \theta {\rm d}
\varphi \nonumber \\
&=& \int_0^\pi \int_0^{2 \pi} \frac{ \rmi Y_{l'm'}^\ast \er \lom
Y_{lm}}{\sqrt{l (l+1) l' (l'+1)}} \sin \theta {\rm d} \theta {\rm d}
\varphi = 0. \label{orthogXeX}
\end{eqnarray}
follows. Below we give some properties of vector spherical
harmonics:
\begin{equation}
\lom {\bf X}_{lm} = \sqrt{l (l+1)} Y_{lm}, \qquad \lom ( \er \times
{\bf X}_{lm})=0. \label{X-properties}
\end{equation}

Scalar (vector) spherical harmonics satisfy the scalar (vector)
equation for eigenfunctions of the squared orbital angular momentum
operator $\lom^2$. The orthogonality condition for the scalar
(\ref{orthogY}) and vector (\ref{orthogX}) spherical harmonics have
the same form. Therefore, one can hope to combine them into one
mathematical object.

\section{Electrodynamic spherical harmonic: definition and properties}

We define an electrodynamic spherical harmonic as a second rank
tensor in three-dimensional space
\begin{equation}
F_{lm}=Y_{lm} \er \otimes \er + {\bf X}_{lm} \otimes \et + (\er
\times {\bf X}_{lm}) \otimes \ep. \label{Flm}
\end{equation}

The first term of (\ref{Flm}) determines the longitudinal part of
the tensor. It is calculated by means of the scalar spherical
function. The last two summands of (\ref{Flm}) fix the transverse
solution, in the plane ($\theta$, $\varphi$) perpendicular to the
direction $\er$. It includes vector spherical harmonics. The left
and right vectors in dyads form two sets of orthogonal vectors:
$(Y_{lm} \er, {\bf X}_{lm}, \er \times {\bf X}_{lm})$ and $(\er,
\et, \ep)$.

\subsection{Orthonormalization}

Multiplying the electrodynamic spherical harmonic by the Hermitian
conjugate $F_{lm}^+$ we get to
\begin{equation}
F_{l'm'}^+ F_{lm}=Y^\ast_{l'm'} Y_{lm} \er \otimes \er + {\bf
X}^\ast_{l'm'} {\bf X}_{lm} (\et \otimes \et + \ep \otimes \ep) +
(\er ({\bf X}_{l'm'}^\ast \times {\bf X}_{lm})) \er^\times.
\label{F2}
\end{equation}

The quantity before each dyad is orthogonal or normalized as
(\ref{orthogY}), (\ref{orthogX}), or (\ref{orthogXeX}). Therefore,
the orthonormalization condition for electrodynamic spherical
harmonics is
\begin{equation}
\int_0^\pi \int_0^{2 \pi} F_{lm}^+(\theta, \varphi) F_{lm}(\theta,
\varphi) \sin \theta {\rm d} \theta {\rm d} \varphi = {\bf 1}
\delta_{l'l} \delta_{m'm}. \label{orthogF}
\end{equation}
Here we consider the vectors $\er$, $\et$, and $\ep$ to be constant
in dyads and dual tensor $\er^\times$. If tensors $F_{lm}$ are the
{\it solutions} of equations, we can always write these solutions in
components. For each component of the tensor $F_{lm}^+ F_{lm}$ the
orthonormalization is carried out. We will obtain the same, if the
dependence of the orts $\er$, $\et$, $\ep$ on angles in (\ref{F2})
is omitted, i.e. they are regarded as constants. However, when
substituting $F_{lm}$ in equations, we should take into account the
angular dependence of the orts.

\subsection{Explicit form}

Let us substitute the explicit expression for the vector operator
$\lom$
\begin{equation}
\lom=-\rmi \ep \frac{\partial}{\partial \theta} + \rmi
\frac{\et}{\sin \theta} \frac{\partial}{\partial \varphi}
\label{L-explicit}
\end{equation}
into equation (\ref{Flm}). Then the electrodynamic spherical
harmonic is equal to
\begin{equation}
F_{lm} = \left[ \er \otimes \er + \frac{\rmi}{\sin \theta \sqrt{l
(l+1)}} \left( I \frac{\partial}{\partial \varphi} - \er^\times \sin
\theta \frac{\partial}{\partial \theta} \right) \right] Y_{lm},
\label{F-explic1}
\end{equation}
where $I=-\er^\times\er^\times={\bf 1} - \er \otimes \er$ is the
projection operator onto the plane ($\theta$, $\varphi$). To
calculate the derivatives one can replace them by means of operators
$L_z$ and $L_\pm=L_x \pm L_y$ as
\[
\frac{\partial}{\partial \varphi} = \rmi L_z, \qquad
\frac{\partial}{\partial \theta} = \frac{1}{2} ( {\rm e}^{-\rmi
\varphi} L_+ - {\rm e}^{\rmi \varphi} L_-),
\]
because their action on the scalar spherical harmonic is well-known:
\begin{equation}
L_+ Y_{lm} = \sqrt{(l-m)(l+m+1)} Y_{l,m+1}, \qquad L_- Y_{lm} =
\sqrt{(l+m)(l-m+1)} Y_{l,m-1}.
\end{equation}

Hence, the electrodynamic spherical harmonic can be presented as
follows
\begin{equation}
F_{lm} = \left[ \er \otimes \er - \frac{1}{\sin \theta \sqrt{l
(l+1)}} \left( I L_z + \er^\times \frac{\rmi \sin \theta}{2} ({\rm
e}^{-\rmi \varphi} L_+ - {\rm e}^{\rmi \varphi} L_-) \right) \right]
Y_{lm}. \label{F-explic2}
\end{equation}

It is easy to exclude the operators from (\ref{F-explic2}). The
final formula for tensor $F_{lm}$ contains scalar spherical
harmonics as angle dependence. Unit vectors $\er$, $\et$, $\ep$
determine the structure of the tensor in three-dimensional space.
$F_{lm}$ is formed by three basic tensors: $\er \otimes \er$,
$\er^\times$, and $I$. Hence, it commutes with each of these
tensors.

\subsection{Invariants}

The first invariant of the electrodynamic spherical harmonic as
tensor quantity is its trace. The trace of the tensor (\ref{Flm})
equals
\begin{equation}
{\rm tr} (F_{lm}) = Y_{lm} + 2 (\et {\bf X}_{lm}). \label{traceF}
\end{equation}

The second invariant is determinant
\begin{equation}
\det (F_{lm}) = Y_{lm} {\bf X}_{lm}^2. \label{detF}
\end{equation}

The third invariant of three-dimensional tensor $F_{lm}$ is the
trace of the adjoint tensor $\overline{F}_{lm}$. Adjoint tensor is
defined by $\overline{F}_{lm} F_{lm}=F_{lm}
\overline{F}_{lm}=\det(F_{lm}) {\bf 1}$ and equals
\begin{equation}
\overline{F}_{lm} = {\bf X}_{lm}^2 \er \otimes \er + Y_{lm} \et
\otimes {\bf X_{lm}} + Y_{lm} \ep \otimes {\er \times \bf X_{lm}}.
\label{adjF}
\end{equation}
Further the trace is easily calculated:
\begin{equation}
{\rm tr} (\overline{F}_{lm}) =  {\bf X}_{lm}^2 + 2 Y_{lm} (\et {\bf
X_{lm}}). \label{tradjF}
\end{equation}
Using these three invariants one can find other ones. For example,
the trace of squared tensor is determined from equation ${\rm tr}
(F_{lm}^2)=({\rm tr} (F_{lm}))^2 - 2 {\rm tr} (\overline{F}_{lm})$.


\subsection{Generalization of electrodynamic spherical harmonic}

The main condition on electrodynamic spherical harmonic is the
orthonormalization (\ref{orthogF}). There is more general form of
the second rank tensor spherical harmonic satisfying this equation.
It is
\begin{equation}
G_{lm}=Y_{lm} \er \otimes {\bf a} + {\bf X}_{lm} \otimes {\bf b} +
(\er \times {\bf X}_{lm}) \otimes {\bf c}. \label{Glm}
\end{equation}
Unit vectors ${\bf a}$, ${\bf b}$, and ${\bf c}$ form the orthogonal
basis in three-dimensional space. In equation (\ref{Flm}) we have
assumed ${\bf a}=\er$, ${\bf b}=\et$, {${\bf c}=\ep$} because of the
spherical symmetry of the function. If we will take ${\bf a}=\er$,
${\bf b}=\et$, {${\bf c}=-\ep$}, then the electrodynamic spherical
harmonic becomes more complex function in explicit form, however its
invariants are simplified (e.g., the trace equals ${\rm tr} (F_{lm})
= Y_{lm}$).

\section{Solution of Maxwell's equations}

In this section we will find the spherically symmetric solutions
(i.e. electric $\ee$ and magnetic $\hh$ fields) of the Maxwell
equations
\begin{equation}
\nabla \times \ee (\rr) = \rmi k \mu \hh (\rr), \qquad \nabla \times
\hh (\rr) = - \rmi k \varepsilon \ee (\rr) \label{Meq}
\end{equation}
for the monochromatic electromagnetic waves in isotropic medium with
dielectric permittivity $\varepsilon$ and magnetic permeability
$\mu$. The quantity $k=\omega/c$ is called wavenumber in vacuum, and
$\omega$ is the wave frequency. Arbitrary time dependence can be
obtained by using the linear superposition of monochromatic waves
$\ee(\rr) \exp(-\rmi \omega t)$.

We will search the solution in the form
\begin{equation}
\ee (\rr) = F_{lm}(\theta, \varphi) \ee^l(r). \label{solut}
\end{equation}
The components of the vector $\ee^l$ in spherical coordinates depend
only on the radial coordinate $r$. The dependence on the angle
coordinates presents only in the basis vectors. So, the vector
$\ee^l$ looks like
\begin{equation}
\ee^l (r) = E^l_r(r) \er + E^l_\theta (r) \et + E^l_\varphi (r) \ep.
\end{equation}

Further we should calculate ${\rm rot} \ee$. By substituting
Hamilton's operator (\ref{nabla2}) one obtains
\begin{equation}
\nabla \times \ee = \er^\times \frac{\partial \ee}{\partial r} -
\frac{\rmi}{r} \lom (\er {\buildrel \downarrow \over \ee}) +
\frac{\rmi}{r} \er (\lom \ee), \label{rotE1}
\end{equation}
where the arrow $\downarrow$ implies that operator $\lom$ acts only
on vector $\ee$, but not $\er$. Let us calculate each summand of
equation (\ref{rotE1}) applying the solution (\ref{solut}). The
first term is of the form
\begin{equation}
\er^\times \frac{\partial \ee}{\partial r}= F_{lm} \er^\times
\frac{\partial \ee^l}{\partial r},
\end{equation}
where the commutation relation $[F_{lm},\er^\times]=0$ is taken into
account. The second summand yields
\begin{equation}
\lom (\er {\buildrel \downarrow \over \ee}) = \lom (\er \ee) - \lom
({\buildrel \downarrow \over \er} \ee) = \lom (Y_{lm} E^l_r) - \lom
({\buildrel \downarrow \over \er} \ee).
\end{equation}
The quantity $\lom ({\buildrel \downarrow \over \er} \ee)$ is easily
calculated using the explicit expression (\ref{L-explicit}) of the
operator $\lom$ and the relationships $\partial \er/ \partial \theta
= \et$ and $\partial \er/ \partial \varphi = \ep \sin\theta$:
\begin{equation}
\lom ({\buildrel \downarrow \over \er} \ee) = - \rmi \er^\times \ee.
\end{equation}
So, we get the formula
\begin{equation}
\lom (\er {\buildrel \downarrow \over \ee}) = F_{lm} \left(
\sqrt{l(l+1)} \et \otimes \er + \rmi \er^\times \right) \ee^l.
\end{equation}
The third term in (\ref{rotE1}) can be rewritten using the equation
(\ref{X-properties}):
\begin{equation}
\er (\lom \ee) = \er \left( E^l_r \lom (Y_{lm} \er) + E^l_\theta
\lom{\bf X}_{lm} + E^l_\varphi \lom(\er \times {\bf X}_{lm}) \right)
= \sqrt{l(l+1)} F_{lm} (\er \otimes \et) \ee^l.
\end{equation}

Finally, the curl of the electric field vector $\ee$ equals
\begin{equation}
\nabla \times \ee = F_{lm} (\theta, \varphi) \left( \er^\times
\frac{\partial}{\partial r} + \frac{1}{r} \er^\times - \frac{\rmi
\sqrt{l(l+1)}}{r} \ep^\times \right) \ee^l (r). \label{rotE2}
\end{equation}

In expression (\ref{rotE2}) we have took into account the
derivatives on the angle variables. Therefore, further the orts of
spherical coordinates $\er$, $\et$, and $\ep$ can be considered as
constants. The Maxwell equations are reduced to the set of ordinary
differential equations
\begin{eqnarray}
\er^\times \frac{{\rm d} \ee^l}{{\rm d} r} + \frac{1}{r} \er^\times
\ee^l - \frac{\rmi \sqrt{l(l+1)}}{r} \ep^\times \ee^l = \rmi k
\mu \hh^l, \nonumber \\
\er^\times \frac{{\rm d} \hh^l}{{\rm d} r} + \frac{1}{r} \er^\times
\hh^l - \frac{\rmi \sqrt{l(l+1)}}{r} \ep^\times \hh^l = - \rmi k
\varepsilon \ee^l. \label{ODE1}
\end{eqnarray}

Equations (\ref{ODE1}) allow to determine the radial dependence of
the fields. Multiplying the set of equations (\ref{ODE1}) by the
unit vector $\er$ we can express the longitudinal components of the
fields as follows
\begin{equation}
E^l_r = -\frac{\sqrt{l(l+1)}}{\varepsilon k r} H^l_\theta, \qquad
H^l_r = \frac{\sqrt{l(l+1)}}{\mu k r} E^l_\theta. \label{longitEH}
\end{equation}
The tangential field components $\ee^l_{\rm t}=I \ee^l= E^l_\theta
\et + E^l_\varphi \ep$ and $\hh^l_{\rm t}=I \hh^l$ are determined
from the equations which can be written in matrix form:
\begin{equation}
\frac{{\rm d} (r \ww)}{{\rm d} r} = i k M (r \ww), \label{ODE2}
\end{equation}
where
\begin{eqnarray}
\ww=\left( \begin{array}{c} \hh^l_{\rm t} \\ \ee^l_{\rm t}
\end{array}  \right), \qquad M= \left( \begin{array}{cc} 0 & \varepsilon A
\\ -\mu A & 0 \end{array}  \right), \qquad A= \er^\times -
\frac{l(l+1)}{\varepsilon \mu k^2 r^2} \ep \otimes \et.
\label{definODE2}
\end{eqnarray}

Equation (\ref{ODE2}) is satisfied for inhomogeneous media
$\varepsilon(r)$, $\mu(r)$, too. Such matrix equation can be solved
numerically for arbitrary medium, or analytically for homogeneous
one. Let us find tangential field components $\ww$ when
$\varepsilon=const$ and $\mu=const$. The simplest way is to write
the equation for projection $W_\theta=(H^l_\theta, E^l_\theta)$:
\begin{equation}
\frac{{\rm d}^2 (r W_\theta)}{{\rm d} r^2} + \left( k^2 \varepsilon
\mu - \frac{l (l+1)}{r^2} \right) (r W_\theta)=0, \label{Wtheta-eq}
\end{equation}
The solutions of the equation (\ref{Wtheta-eq}) are well-known and
can be presented as
\begin{equation}
W_\theta = f^{(1)} \left( \begin{array}{c} c_1 \\ c'_1
\end{array}  \right)  + f^{(2)} \left( \begin{array}{c} c_2 \\ c'_2
\end{array}  \right) = (f^{(1)} {\bf c}_1 + f^{(2)} {\bf c}_2)
\left( \begin{array}{c} \et \\ \ep \end{array}  \right),
\end{equation}
where ${\bf c}_1$ and ${\bf c}_2$ are constant vectors. The couples
of independent solutions are spherical Bessel functions $f^{(1)}=j_l
(n k r)$, $f^{(2)}=j_{-l-1} (n k r)$ or spherical Hankel functions
of the first and second kind $f^{(1)} = h_l^{(1)} (n k r)$, $f^{(2)}
= h_{l}^{(2)} (n k r)$. $n=\sqrt{\varepsilon \mu}$ is the refractive
index. After determining the $\varphi$-projections of the fields as
\begin{equation}
W_\varphi = \frac{\rmi}{k r \varepsilon \mu} \left(
\begin{array}{cc} 0 & -\varepsilon \\ \mu & 0 \end{array}  \right)
\frac{{\rm d} (r W_\theta)}{{\rm d} r} \label{Wphi}
\end{equation}
one can write the transverse vector field
\begin{equation}
\ww(r)=\left( \begin{array}{cc} \eta_1(r) & \eta_2(r)
\\ \zeta_1(r) & \zeta_2(r) \end{array} \right) \left(
\begin{array}{c} {\bf c}_1 \\ {\bf c}_2 \end{array} \right),
\label{WSC}
\end{equation}
where
\begin{eqnarray}
\eta_{1,2} = f^{(1,2)} \et \otimes \et - \frac{\rmi }{\mu k r}
\frac{{\rm d} (r f^{(1,2)})}{{\rm d} r} \ep \otimes \ep, \nonumber \\
\zeta_{1,2} = f^{(1,2)} \et \otimes \ep + \frac{\rmi}{\varepsilon k
r} \frac{{\rm d}( r f^{(1,2)})}{{\rm d} r} \ep \otimes \et.
\label{eta_zeta_isotr}
\end{eqnarray}

Tangential field vector $\ww$ plays an important part, because it is
continuous on the spherical surface. That is why it can be applied
for the study of electromagnetic wave diffraction by a sphere.

\section{Conclusion}

Thus, the general solution of the Maxwell equations is of the form
\begin{equation}
\left( \begin{array}{cc} \hh (\rr)
\\ \ee(\rr) \end{array} \right)= \sum_{l=0}^\infty \sum_{m=-l}^l F_{lm} (\theta, \varphi) V^l(r)
\left( \begin{array}{cc} \eta^l_1(r) & \eta^l_2(r)
\\ \zeta^l_1(r) & \zeta^l_2(r) \end{array} \right) \left(
\begin{array}{c} {\bf c}^l_1 \\ {\bf c}^l_2 \end{array} \right),
\label{gen_sol}
\end{equation}
where $V^l$ is the matrix that takes into account the longitudinal
components of electric and magnetic fields. This matrix can be
easily calculated from the equation (\ref{longitEH}). In each
partial solution included into the general one (\ref{gen_sol}) the
radial and angle variables are separated. Using the
orthonormalization (\ref{orthogF}) for the electrodynamic spherical
harmonic $F_{lm}$, each partial wave can be easily singled out:
\begin{equation}
V^l(r) \left( \begin{array}{cc} \eta^l_1(r) & \eta^l_2(r) \\
\zeta^l_1(r) & \zeta^l_2(r) \end{array} \right) \left(
\begin{array}{c} {\bf c}^l_1 \\ {\bf c}^l_2 \end{array} \right) =
\int_0^\pi \int_0^{2 \pi} F^+_{lm} (\theta', \varphi') \left(
\begin{array}{cc} \hh (r, \theta', \varphi') \\ \ee(r, \theta',
\varphi') \end{array} \right) \sin \theta' {\rm d} \theta' {\rm d}
\varphi'.
\end{equation}
This property of the electrodynamic spherical harmonic is very
useful for the investigation of electromagnetic wave scattering. In
scattering, the boundary condition is the single equation for
tangential fields $\ww$. It is easily solved, if the
orthonormalization is applied. Some attempts of investigation of
scattering in the similar manner as described above have been made
in \cite{Novitsky06jpa_sph}.

In the general solution (\ref{gen_sol}) the constants ${\bf
c}_{1,2}$ determined by initial conditions are explicitly shown.
Vectors ${\bf c}_{1}$ and ${\bf c}_{2}$ set independent solutions.
For instance, if ${\bf c}_{1}=0$, then the radial dependence is
determined by the function $f^{(2)}(r)$, and vice versa. If ${\bf
c}_{2}=0$ and $f^{(1)}(r)=h^{(1)}(n k r)$, then the field
({\ref{gen_sol}) is the multipole expansion \cite{Jackson}. The
amplitude of electric multipole field is equal to $a_E (l,m)={\b
c}_1 \et$, the amplitude of magnetic multipole field is equal to
$a_M (l,m)={\bf c}_1 \ep$. So, vector ${\bf c}_1$ can be called
vector amplitude of multipole fields.

In further investigations we will study the scattering and multipole
expansion of electromagnetic fields in details.

\end{document}